\documentclass[prl,aps,twocolumn]{revtex4}

\usepackage{times}
\usepackage{graphicx}
\usepackage{float}
\usepackage{latexsym,amsmath,amssymb,bm,euscript}
\usepackage{color}
\usepackage{subfigure}
\usepackage{epstopdf}
\usepackage[colorlinks=true,linkcolor=blue,citecolor=blue]{hyperref}
\usepackage{type1cm}

\usepackage{appendix}

\usepackage{extarrows}
\usepackage{slashed}
\usepackage{lipsum}
\usepackage{srcltx}
\usepackage{mathtools}

\usepackage{tikz}
\usepackage{tikz-cd}
\usetikzlibrary{arrows}
\usetikzlibrary{intersections}
\usetikzlibrary{shapes.geometric}
\usetikzlibrary{shapes, positioning}

\def\be{\begin{equation}}
\def\ee{\end{equation}}

\begin{document}
\title{Entanglement Hamiltonian of Many-body Dynamics in Strongly-correlated Systems} 

\author{W. Zhu$^{1}$,  Zhoushen Huang$^2$, Yin-Chen He$^3$, and  Xueda Wen$^4$}
\affiliation{$^1$Westlake Institute of Advanced Study, Westlake University, Hangzhou, 310024, China}
\affiliation{$^2$Argonne National Laboratory, Lemont, IL 60439, USA}
\affiliation{$^3$Perimeter Institute for Theoretical Physics, Waterloo, Ontario N2L 2Y5, Canada }
\affiliation{$^4$Department of Physics, Massachusetts Institute of Technology, Cambridge, MA 02139, USA }

\begin{abstract}
A powerful perspective in understanding non-equilibrium quantum
dynamics is through the time evolution of its entanglement
content. Yet apart from a few guiding principles for the entanglement entropy, 
to date, not much else is known about the refined characters of entanglement propagation.
Here, we unveil signatures of the
entanglement evolving and information propagation out-of-equilibrium, 
from the view of entanglement Hamiltonian.
As a prototypical example, we study quantum quench dynamics of a one-dimensional Bose-Hubbard model
by means of time-dependent density-matrix renormalization group simulation.  
Before reaching equilibration, it is found that a current operator emerges in entanglement Hamiltonian,
implying that entanglement spreading is carried by particle flow. 
In the long-time limit subsystem enters a steady phase, evidenced by 
the dynamic convergence of the entanglement Hamiltonian to the expectation of a thermal ensemble.
Importantly, entanglement temperature of steady state is spatially independent, 
which provides an intuitive trait of equilibrium.
We demonstrate that these features are consistent with predictions from conformal field theory.
These findings not only provide crucial information on how equilibrium statistical mechanics 
emerges in many-body dynamics, 
but also add a tool to exploring quantum dynamics from perspective of entanglement Hamiltonian. 
\end{abstract}


\maketitle


\textit{Introduction.---} The power of classical statistical mechanics is rooted 
in the ergodic hypothesis, but in closed quantum many-body systems, 
how ``memories'' are forgotten in a realistic time scale \cite{Luca2016,Mori2018,BORGONOVI2016,Gogolin2016}
--- how steady states and thermal behavior at later times emerge
dynamically \cite{Deutsch1991,Srednicki1994,Rigol2008}--- 
remains an actively investigated topic \cite{Huse2015,Deutsch2018,Abanin2019,Anatoli2011} .
Recently, there is a surge of theoretical interests on the problems of non-equilibrium quantum dynamics, 
thanks in part to significant progress in experimental techniques that has made the
dynamics of quantum systems accessible
\cite{Eisert2015,Trotzky2012,Roos2014,Monroe2014,Neill2016,Lukin2017a,Lukin2017b,Monroe2017,Greiner2016,Greiner2018,Tang2018}. 
In many cases,
particularly in interacting systems, however, to directly access 
such dynamics remains technically challenging due to the increasing
amount of correlations generated over time \cite{Lieb1972,Nachtergaele2006}.

From an entanglement point of view, these correlations are a
consequence of entangled quasiparticle pairs being constantly
generated and propagating into different parts of the system
\cite{Lieb1972,Nachtergaele2006,Cardy2005,Cardy2006,Calabrese_2016,Hastings2010}. The
dynamics of these quasiparticles have been shown to reflect the
underlying nature of their hosting systems, e.g., ballistic in
thermalizing systems \cite{Cardy2005,Calabrese2006,Huse2013} versus logarithmic in localized systems
\cite{Bardarson2012,Iglo2012,Burrell2013,Abanin2013}. In
many of these examples, propagation of entanglement also spreads
conserved quantities which can serve as information carrier
\cite{Lieb1972,Qi2018,Alba2017,Nahum2017}. An important aspect to
understanding quantum dynamics and the emergence of equilibration is
therefore to understand the dynamics of quantum entanglement \cite{Abanin2019},
even in systems without identifiable quasiparticle content \cite{Huse2013,Lauchli2008,Pal2018,HongLiu2014,Casini2016}.
In this context, entanglement dynamics is also connected with information loss and scrambling
\cite{Hosur2016,Swingle2016,ZWLiu2018,Keyserlingk2018,Zhui2017,Asplund2015}.

In equilibrium condensed matter systems, entanglement-based analysis has already proved to be a profitable tool
as a diagnostic of strong correlations, from the presence of topological order to the onset of
quantum criticality \cite{Laflorencie2016}.  Indeed, the scaling of
entanglement entropy characterizes the quantum statistics of
quasiparticles \cite{Levin2006,Kitaev2006}, and entanglement spectrum
holds a direct relation between bulk and edge physics
\cite{Haldane2008}, both of which highlight the wealth of information
encoded in entanglement.  While entanglement entropy and entanglement
spectrum are important measures of quantum information, entanglement
Hamiltonian (EH) is a more fundamental object.  The EH is a sum of
local ``energy'' density $\mathcal{H}(x)$ weighted by a local entanglement
temperature $\beta(x)$: $H_E= \int dx \beta(x) \mathcal{H}(x)$.
The relationship between EH and reduced density matrix of a subsystem
(A), $\rho_A=e^{-H_E}$, implies that $\rho_A$ can be interpreted as a
canonical ensemble with energy density $\mathcal{H}(x)$ in local thermal
equilibrium at temperature $\beta^{-1}(x)$.  Therefore, knowledge of the EH
could offer an alternative picture of how subsystem A behaves by
appealing to our intuition of thermodynamics.  However, even for
static systems, precise knowledge about their EH is rare.  The only
exact result of EH known to date pertains to
integrable systems described by (1+1)-dimensional conformal field
theory (CFT) \cite{Bisognano1975,SRyu2016}, for which the local
temperature $\beta(x)$ satisfies a spatially arch-like envelope
function.  Recently, numerical efforts have attempted to
obtain the EH in static interacting systems using various methods
\cite{Assaad2018,Dalmonte2018,WZhu2018}, and have shed some light on
this technically challenging problem.  As for time-evolving systems,
although results for non-interacting cases have been obtained
\cite{Tonni2016,Xueda2018}, 
the quantitative role of EH in strongly-correlated systems remains unexplored,
and it is far from obvious how the time dependence of EH should be.

In this work, we study the EH in the
quench dynamics of Bose Hubbard model, a prototypical non-integrable
system, based on time-dependent density-matrix renormalization group
(t-DMRG) approach \cite{White1992,White2004}.  With the help of a
recently developed numerical scheme \cite{WZhu2018}, we are able to
track the time dependence of the EH in real time.  Our main findings
are that: 1) a current operator emerges in the EH before the system
reaches equilibration, reflecting the propagation of entanglement
carried by particle flow; 2) in the long-time limit, the EH becomes
nearly stationary and demonstrates features of equilibration; 3) the
long-time steady state exhibits a spatially independent
entanglement temperature, signaling the subsystem becomes locally thermal.
All above results are endorsed by CFT.  
These findings imply that the EH can be used to effectively
investigate the emergence of subsystem equilibration under the unitary dynamics of the full system, 
which sets up a valuable paradigm for exploring entanglement dynamics out-of-equilibrium.

\textit{Preliminary.---}
We begin by discussing the salient features of the EH dynamics 
after a quantum quench, in the framework of 1+1D CFT. 
We consider a 1D chain with finite length $L$ defined on $x \in [0, \, L]$, 
and the subsystem $A$ under consideration is chosen as $[0, l]$.
At time $t = 0$, we start from an initial state with short-range 
entanglement, which may be considered as the ground state of a gapped Hamiltonian.
At $t>0$ we evolve it with a CFT Hamiltonian $H_{\text{CFT}}=\int dx \mathcal{H}(x)$.
We consider the case where the time scale $t$ is smaller than the total length $L$, 
such that the other boundary at $x=L$ can be safely neglected.

Based on conformal mappings, we obtained the exact form of the EH (See supplementary materials for details \cite{sm}). 
Importantly, we found that in the long-time limit, the EH of subsystem A is the sum of
$\mathcal{H}(x)$ weighted by a spatially dependent finite temperature $\beta^{-1}(x)$,
indicating that the reduced density matrix $\rho_A(t)$ takes the form of 
a thermal ensemble.
To be specific, 
in the long time limit $t\gg l$, one obtains the EH $H_E=\int dx \beta(x) \mathcal{H}(x)$, 
with the envelope function \cite{sm}
\begin{align}
&\beta(x)= 2\beta_0 \cdot\frac{\sinh(\pi(l+x)/\beta_0)\sinh(\pi(l-x)/\beta_0)}{\sinh(2\pi l/\beta_0)}, (t\gg l). \label{eq:cft2}
\end{align}
Here $\beta_0$ characterizes the correlation length of the gapped pre-quench state \cite{Calabrese_2016},
and it also qualifies the effective ``temperature'' of energy density of the system using pre-quench state \cite{sm}.
In addition, as notable byproducts, CFT also gives time dependence of
entanglement entropy to the leading order
\cite{Cardy2005,Cardy2006,sm}:  
\begin{eqnarray}
S(t)=\begin{cases}
\frac{3c}{\pi \beta_0}t,\,\,\,\,\,\quad t<l\\
\frac{3c}{\pi \beta_0}l, \,\,\,\,\,\quad t>l   
\end{cases}, \label{eq:cft_ee}
\end{eqnarray}
where $c$ is the central charge of the underlying CFT.  That is, the
entanglement entropy grows linearly in time until it saturates at a
value satisfying the volume law \cite{sm}.

\begin{figure}[b]
	\includegraphics[width=0.45\textwidth]{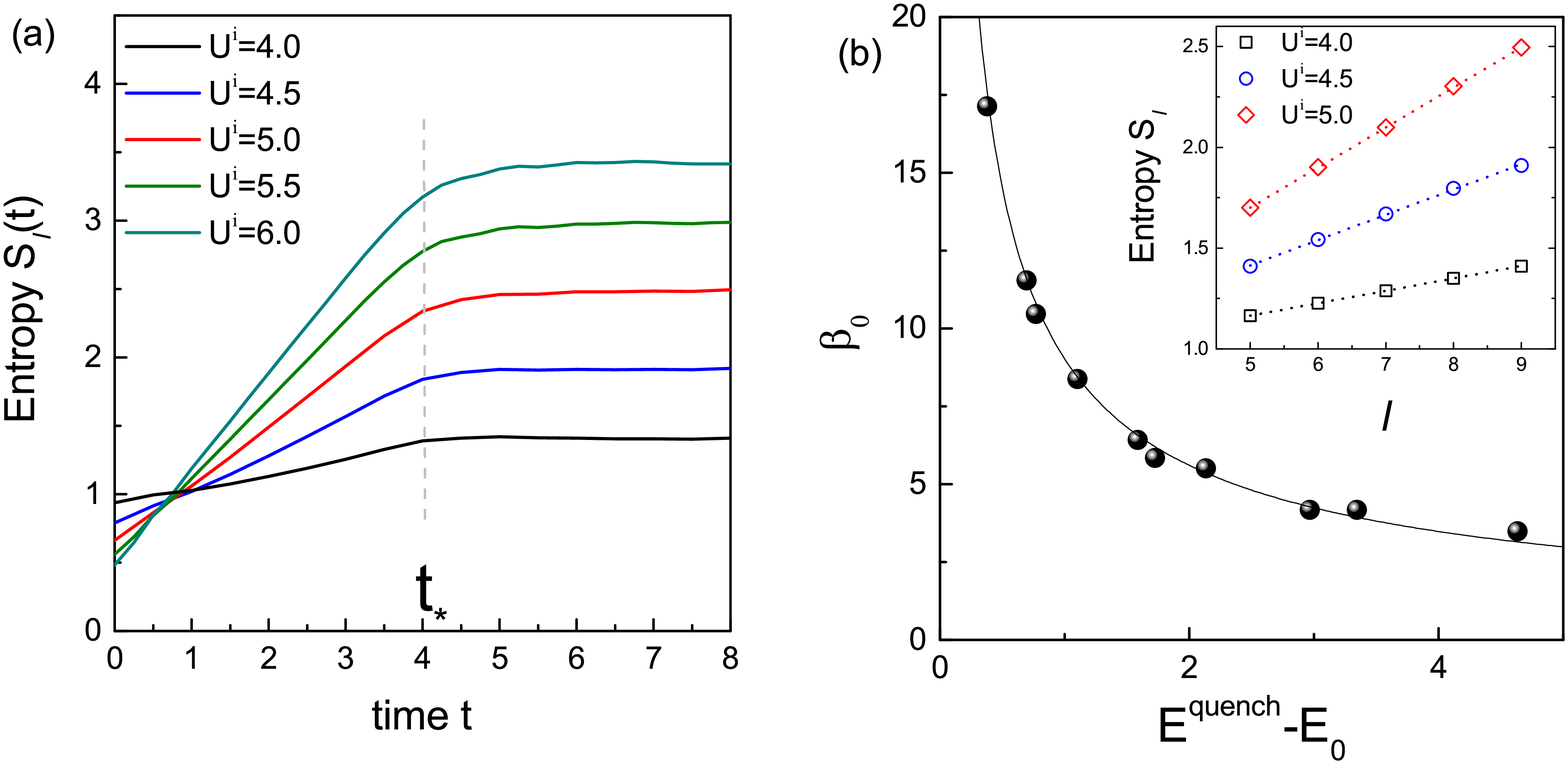}
	\caption{\textbf{Dynamics of the entanglement entropy.}
		(a) Time-evolution of entanglement entropy by quenching from various $U^{\mathbf{i}}$ to $U^{\mathbf{f}}=3.3$.
		(b) Effective temperature $\beta_0$ as a function of $E^{\mathbf{quench}}-E_0$, 
		where $E_0$ is the lowest energy of post-quench Hamiltonian $\hat H(U^{\mathbf f})$ and
		$E^{\mathbf{quench}}= \langle \Psi(t=0)|H(U^{\mathbf f})|\Psi(t=0) \rangle$. 
		The black line is the best fit to $\beta_0\propto (E^{\mathbf{quench}}-E_0)^\alpha,\alpha=-0.641 \pm 0.012$.
		Inset: Linear scaling of $S_{\ell}=\frac{\pi c}{3\beta_0}\ell$ to the length of the subsystem $\ell$.
	} \label{fig:EE_BH}
\end{figure}

\begin{figure*}[t]
	\includegraphics[width=0.95\textwidth]{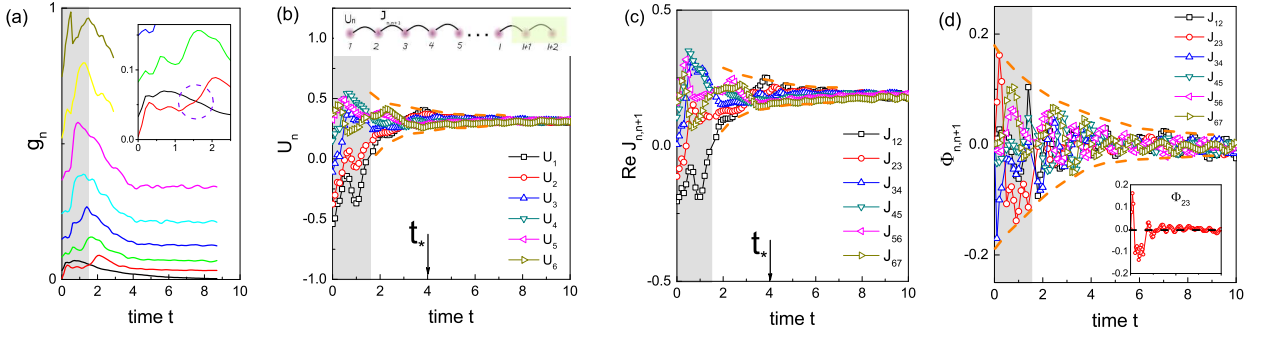}
	\caption{\textbf{Dynamics of the EH.}
		(a) Spectrum of correlation matrix $G_{ab}(t)$. 
		The lowest and second lowest eigenvalue crosses with each other at $t_0\approx 1.65$ (inset).
		The shaded area shows the short time regime $t<t_0$.  
		The parameters of the EH (see Eq. \ref{eq:EH}) as a function of time: (b) interaction strength $U_n(t)$,
		(c) real part of couplings $Re J_{n,n+1}(t)$, (d) relative phase of couplings $\Phi_{n,n+1}(t)=\arg J_{n,n+1}$, 
		where $n$ labels spatial lattice sites. 
		Here we quench the Bose-Hubbard model (Eq. \ref{eq:BH}) from $U^{\mathbf{i}}=5.0$ to $U^{\mathbf{f}}=3.3$.
		The total system size $L=48$ and the typical subsystem length is $\ell=9$. 
		Different symbols label local coupling and interaction strengths.
		The brown dashed line is guide to eye.
		Inset of (b) is the cartoon picture of one dimension chain and entanglement bipartition.
	} \label{fig:EH_BH}
\end{figure*}

\textit{Model and Method.---}
We now turn to a paradigmatic
non-integrable model, the one-dimensional Bose-Hubbard model, which
has been experimentally realized with ultracold gases in deep
optical lattices \cite{Cazalilla2011},
\begin{eqnarray}\label{eq:BH}
\hat H= -J\sum_{i} (b^\dagger_i b_{i+1} + h.c.) + \frac{U}{2}\sum_i n_i(n_i-1),
\end{eqnarray}
where $b^\dagger_i (b_i)$ is the boson creation (annihilation)
operator and $n_j=b^\dagger_i b_i$ is the on-site density operator.
Throughout this work, we consider a uniform Hamiltonian density, i.e.
the physical coupling $J$ (set to $J=1$) and interaction $U$ are
spatially independent.  In the equilibrium case, at fixed filling
$\langle n_i\rangle=1$, a critical value $U_c\approx 3.38$
\cite{Fisher1989,Kuhner1998} separates a Mott insulating phase
($U>U_c$) from a superfluid phase ($U<U_c$), the latter described by
an effective Luttinger liquid theory with $c=1$.  Below we set the
initial state in the Mott phase as the ground state of
$H$ with pre-quench condition $U^{\mathbf i} > U_c$, 
and investigate its quench dynamics under
the $H$ with post-quench condition $U^{\mathbf f} < U_c$.

\newcommand{\uu}{\mathcal{U}}

To simulate the unitary time evolution
$|\Psi(t)\rangle = \uu(t)|\Psi(t=0)\rangle$, we use the time-dependent
density-matrix renormalization group (t-DMRG)
\cite{White1992,White2004}.  We apply a second-order Trotter
decomposition of the short time propagator
$\uu(\Delta t)=\exp(-i \Delta t \hat H)$ into a product of term which
acts only on two nearest-neighbor sites. We use a dimension up to
$5120$, which guarantees that the neglected weight in the Schmidt
decomposition in each time step is less than $10^{-6}$.  Once the
$|\Psi(t)\rangle$ is computed, we partition the one-dimensional chain
of length $L$ into two segments, $\ell$ and $L-\ell$, and calculate
the subsystem reduced density matrix,
$\rho_{\ell}(t)=Tr_{L-\ell} |\Psi(t)\rangle \langle \Psi(t)|$.
The entanglement Hamiltonian is formally defined as
$\rho_A(t)=\exp(-\hat H_E)$, but it is technically challenging to
extract $\hat H_E$ through this definition because the transformation
$\hat H_E(t)=-\ln \rho_A(t)$ is non-linear.  Very recently, a generic
scheme to obtain the operator form of EH has been proposed in
Ref. \cite{WZhu2018}, which we briefly outline here.  The starting
point is to define a set of basis operators $\hat L_a$, which we take
as the boson hopping operator $b^\dagger_i b_j$ and density
interaction operator $n_i(n_i-1)$ according to the form of the
physical Hamiltonian.  These operators define the variational space in
which we search for the ``best'' EH in the form
$H_E = \sum_a w_a \hat L_a$, where $w_a$ are parameters coupled to operators $\hat L_a$. 
Practically, the variational scheme is equivalent to
solve the eigenvalue problem of the correlation matrix 
$G_{ab}=\langle \xi| \hat{L}_a \hat{L}_b |\xi \rangle- \langle \xi|\hat{L}_a |\xi\rangle \langle \xi|\hat{L}_b |\xi\rangle$ \cite{XLQi2017,WZhu2018},
where $|\xi\rangle$ is a reference state chosen here as one eigenstate of $\rho_A$.
The lowest eigenvalue of $G_{ab}$, i.e. $g_0$, 
minimizes the variance $\langle \xi | H_E^2 | \xi\rangle - \langle \xi|H_E|\xi\rangle^2$, 
which can be interpreted as the ``fluctuation'' of ``Hamiltonian'' $H_E =\sum_a w_a L_a$ under $|\xi\rangle$. 
The eigenvector of $g_0$ gives rise to the estimate of $\{w_a\}$. 
It has been confirmed that \cite{WZhu2018}, in the static case
this numerical receipt can give reliable EH that faithfully captures
all features of the reduced density matrices.
In this work, we generalize and formulate this scheme using
matrix-product state ansatz, which is amenable to simulating the time
evolution of the EH within the t-DMRG approach, and works well for
larger system sizes compared to exact diagonalization.

\textit{Entanglement entropy.---} 
We compute the time-dependent entanglement entropy and compare with the CFT results obtained earlier.
Fig. \ref{fig:EE_BH}(a) shows the time evolution of the entanglement entropy for various initial conditions $U^{\mathbf i}$.
For all cases, $S_{\ell}(t)$ shows two temporal regimes:
At short times $t<t_*$, the entropy shows a linear rise,
until it bends over to an almost flat plateau.
The linear increase can be accounted for by the ``ballistic'' propagation of entanglement.
At long times $t>t_*$, the entropy saturates to its steady-state value.
As shown in   inset of Fig. \ref{fig:EE_BH}(b),
the saturation of the entropy depends linearly on the block length,
which clearly exhibits a ``volume-law'' scaling.
In particular, based on the relationship of Eq. \ref{eq:cft_ee}, 
we can extract the pre-quench entanglement temperature $\beta_0$ (or correlation length of the initial state). 
In Fig. \ref{fig:EE_BH}(b), 
we show the dependence of the effective entanglement temperature $\beta_0$ on the post-quench energy above the ground state, $E^{\mathbf{quench}}-E_0$,
where $E^{{\mathbf{quench}}}$ is the energy of the pre-quench state in the post-quench Hamiltonian, and $E_0$ is the post-quench ground state energy.
It is clear that $\beta_0$ monotonically decreases with $E^{\mathbf{quench}}-E_0$.
Our best fitting gives the scaling $\beta_0\propto (E^{\mathbf{quench}}-E_0)^\alpha,\alpha\approx -0.641 \pm 0.012$.
It reflects that a higher initial energy translates to a higher effective temperature.

\textit{Entanglement Hamiltonian.---}
Next we turn to discuss the time evolution of EH. 
Here we assume the EH has following form (detailed discussion see \cite{sm}):
\begin{equation}\label{eq:EH}
H_E(t)=   -\sum_{i}  (J_{i,i+1}(t)b^\dagger_i b_{i+1} + h.c.) +\sum_i  \frac{U_i(t)}{2}n_i(n_i-1).
\end{equation}
We map out the EH at each time step by using the scheme described in the method section \cite{WZhu2018}.
Fig. \ref{fig:EH_BH}(a) shows the spectrum of correlation matrix as a function of time. 
Interestingly, it is found a level crossing between the lowest and 
second lowest eigenvalue around $t_0\approx 1.65$ (inset of Fig. \ref{fig:EH_BH}). 
After this critical time,
the lowest eigenvalue $g_0$ monotonically decreases, implies the trial EH
works better in the time regime $t>t_0$. Next we will focus on the $t>t_0$ regime
and discuss the salient features of the EH.

Fig. \ref{fig:EH_BH}(b-c) shows the time evolution of the interaction strength $U_i(t)$, 
real part of coupling strength $Re J_{i,i+1}(t)$ after a global quench.
First, both $J$ and $U$ show sizable oscillations at early times  $t<t_0$, and later  
the subsequent dynamics gradually reduce (as indicated by the envelope dashed curve). 
In particular, at the long-time limit $t>t_*$, 
all coupling strengths approach almost stationary values. 
Physically, this suggests the subsystem has equilibrated to a steady state.
 
Second, before reaching equilibration, it is found
the imaginary part of boson hopping strength is nonzero. 
To show this, we define the phase angle $\Phi_{i,i+1}=\arg J_{i,i+1} =\tan^{-1}\frac{Im J_{i,i+1}}{Re J_{i,i+1}}$, 
and the phase angle directly relates to the imaginary part of coupling strength  $Im J_{i,i+1}(t)= |J_{i,i+1}|\sin \Phi_{i,i+1}$.
In Fig. \ref{fig:EH_BH}(d), $\Phi_{i,i+1}(t>0)$ shows oscillation behaviors due to the non-equilibrium dynamics. 
For comparison, in the static case we have $\Phi_{i,i+1}(t=0)= 0$. 
Since $Im J_{i,i+1}$ is directly coupled to the current operator $\hat J_c=i[H,x]=i\sum_i (b^\dagger_n b_{n+1}-b_n b^\dagger_{n+1})$ (we set $e=\hbar=1$), 
this implies that time-reversal symmetry is broken, and a non-vanishing particle current flow emerges in time evolution. 
The emergent current flow reflects quasiparticle propagation, which is consistent with 
the picture that quasiparticles serve as entanglement information carriers\cite{Cardy2005}.
The inset of Fig. \ref{fig:EH_BH}(d) single out one typical evolution ($\Phi_{2,3}$). 
It signals that the current first flows from the entanglement cut into the bulk ($\Phi_{2,3}> 0$), 
and then reverse direction ($\Phi_{2,3}<0$), and reduces to zero in the long time. 
This again shows the transport of quasiparticles. 
At long times, the imaginary part tends to vanish with only small fluctuations around zero, 
suggesting that the subsystem has reached equilibrium and net particle flow is absent.
The appearance of current in the EH allows us to conclude that 
information spreading originates in the propagation of quasiparticles 
between the two bipartition constituents \cite{Cardy2005}.

\begin{figure}
	\includegraphics[width=0.20\textwidth]{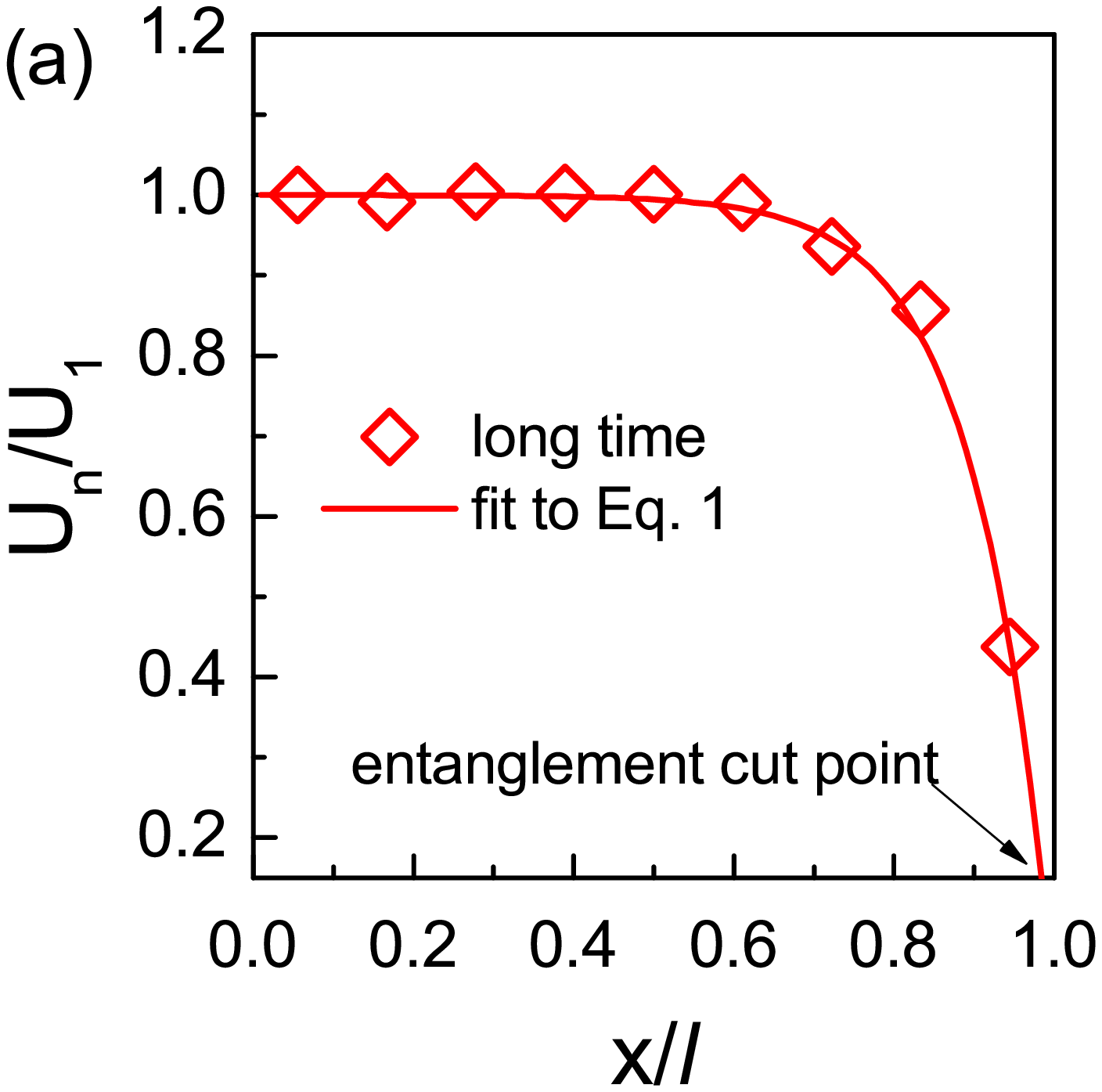}	
	\includegraphics[width=0.27\textwidth]{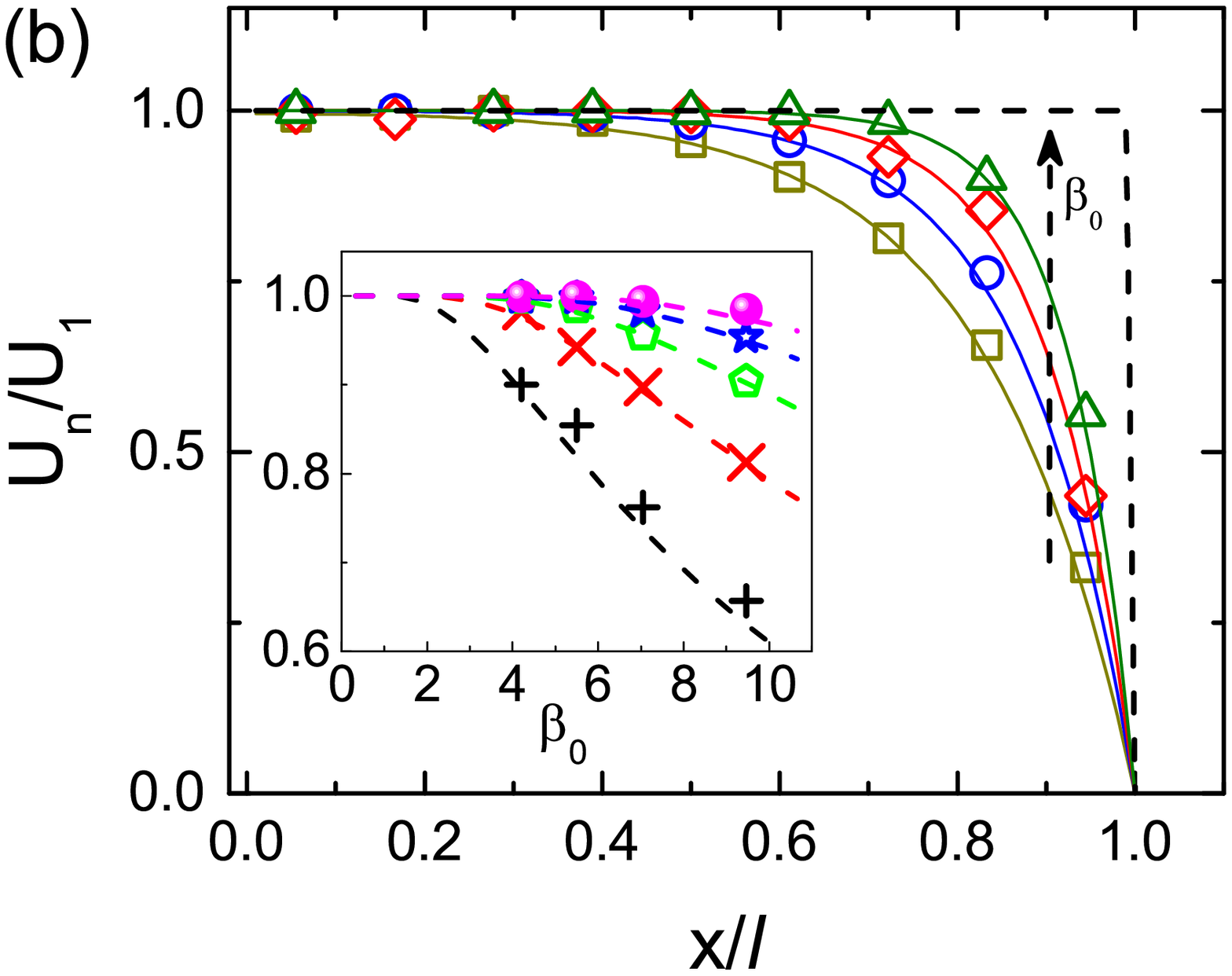}
	\caption{\textbf{Spatial dependence of the EH.}
	(a) Local coupling strengths at long time limit (red diamonds).
	The red line fits Eq. (\ref{eq:cft2}). (b)
	Spatial dependence of local coupling strengths for various quenching parameters:
	$U^{\mathbf i}=4.0,U^{\mathbf f}=3.3$ (black squares), 
	$U^{\mathbf i}=4.5,U^{\mathbf f}=3.3$ (blue circles), 
	$U^{\mathbf i}=5.0,U^{\mathbf f}=3.3$ (red diamonds) and 
	$U^{\mathbf i}=5.5,U^{\mathbf f}=3.3$ (green triangles). 
	The solid lines show best fit to the envelope function Eq. (\ref{eq:cft2}) 
	with various pre-quench temperature $\beta_0$.
	Inset: Interaction strength scaling to infinite temperature.
	} \label{fig:EH_BH_scaling}
\end{figure}

Third, as shown in  Fig. \ref{fig:EH_BH}(b-c), 
at the long time limit $t>t_*$ the evolution of local coupling and interaction strengths at different spatial locations tend to converge to the same value,
indicating that the EH is spatially uniform away from the entanglement cut. 
To further study the spatial dependence of the EH at the long-time limit,
we plot the time-averaged local coupling strengths as a function of distance to the cut in Fig. \ref{fig:EH_BH_scaling}.
In Fig. \ref{fig:EH_BH_scaling}(a), we show the spatial dependence of local interaction strength $U_n/U_1$ at the long times.
In particular, local strengths in the long time limit are nearly uniformly distributed away from the entanglement cut ($x\ll\ell$).
Crucially, this spatial dependence shows excellent agreement with the CFT prediction Eq. (\ref{eq:cft2}).
Moreover, we demonstrate that the residual fluctuations near the entanglement cut $x\sim \ell$
can be interpreted as a finite temperature effect. 
In Fig. \ref{fig:EH_BH_scaling}(b), we show that 
by increasing temperature (through changing quenching parameters as discussed in Fig. \ref{fig:EE_BH}(b)), 
the spatial independence of local strengths becomes sharper near the entanglement cut $x\sim \ell$. 
The consistency with the CFT Eq. (\ref{eq:cft2}) indicates that 
local strengths should be completely flat (shown by dashed line) at infinite temperature,
which is also supported by our numerical results (inset of Fig. \ref{fig:EH_BH_scaling}(b)). 
Physically, spatial dependence of local coupling strengths 
in the EH can be interpreted as a local entanglement temperature $\beta^{-1}(x)$,
and $\rho_A=\exp(-\int dx\beta(x)H_E(x))$
resembles a physical system equilibrated at local temperature $\beta^{-1}(x)$ depending on distance from a ``heat source'' that is subsystem B.
From this point of view, it is appealing that spatially independent $\beta(x)$ 
reveals local temperature reaches the equilibration.

\textit{Summary and Discussion.---}
We have addressed the out-of-equilibrium dynamics of strongly-correlated systems 
from the point of view of entanglement Hamiltonian.
By tracking the time evolution of the entanglement Hamiltonian, we were able to gain remarkable signatures 
of the entanglement propagation and information scrambling.
We demonstrate that, the entanglement Hamiltonian involves an emergent current operator,
which drives the quasiparticle propagation towards equilibration.
In the long-time limit the entanglement Hamiltonian becomes stationary. 
In particular, spatially distributed entanglement temperature satisfies a universal feature
as proposed by the conformal field theory, 
indicating the subsystem indeed reach equilibrium away from the entanglement cut.
Our results shows that entanglement Hamiltonian provides fundamental insight into 
the non-equilibrium dynamics of quantum many-body systems.

In closing, we would like to make several remarks. 
Although the limited system sizes prevent comparison over 
a large range of subsystem sizes, we confirm 
the characters of entanglement Hamiltonian with underlying scaling behavior 
are robust on all of system sizes we can reach \cite{sm}. 
Moreover, we investigate numerically a variety of one-dimensional systems of different kinds \cite{sm}.
Through these studies, our results have implications well beyond the specific model.
Lastly, our findings open up several avenues for future investigation.
For instance, applying these tools for characterizing the presence of
equilibration could be powerful in studying many-body localization \cite{Huse2015,Alet2018,Abanin2019}, 
where one of the key features is the suppression of entanglement.
In addition, taking into account the recent proposal in synthetic quantum systems \cite{Zoller2018},
the dynamics of constructed entanglement Hamiltonian may be valuable for future experiments.






\textit{Note Added---}
At the final stage of preparing this
manuscript, we became aware of a work on entanglement Hamiltonian in non-interacting systems \cite{Tonni2019}.

\textit{Acknowledgments.---}
W.Z. thanks Beni Yoshida for fruitful discussion.
This work was supported by the start-up funding at Westlake University.
Work at Argonne was supported by ANL LDRD Proj. 1007112. 
X.W. is supported by the Gordon and Betty Moore Foundation’s EPiQS initiative through Grant No.GBMF4303 at MIT.
Research at Perimeter Institute (YCH) is supported by the Government of Canada through the Department of Innovation, Science and Economic Development Canada and by the Province of Ontario through the Ministry of Research, Innovation and Science.

\bibliographystyle{apsrev}
\bibliography{time_evolution}{}

\clearpage
\begin{widetext}

\newpage
\section{Entanglement Hamiltonian evolution after a global quantum quench}

In this appendix, we derive the time evolution of entanglement Hamiltonian after a global quantum quench in a (1+1) dimensional conformal field theory (CFT)
\cite{Xueda2018}.
We start from a short-range entangled state $|\phi_0\rangle$, which may be considered as the ground
state of certain gapped Hamiltonians. Then at $t=0$,  $|\phi_0\rangle$ is evolved under a gapless Hamiltonian whose low energy dynamics
can be described by a CFT Hamiltonian $H_{\text{CFT}}$. That is, the time dependent wavefunction is $|\psi(t)\rangle=e^{-iH_{\text{CFT}}t} |\phi_0\rangle$.

The system studied in the main text is of a finite length $L$ defined on $[0, \, L]$, with the subsystem $A$ chosen in the interval $[0, l]$.
We are interested in the case that the time scale $t$ is smaller than the total length $L$ (velocity is set to be $1$), such that the other boundary at $x=L$ may be 
safely neglected. That is, the only two relevant scales in this problem are the subsystem length $l$ and the correlation length in the initial state which we will
introduce shortly. Then the problem is reduced to a global quantum quench in a semi-infinite system $[0,\infty)$ with the subsystem in $[0,l]$.

More explicitly, the initial state we consider has the form $|\phi_0\rangle=e^{-\frac{\beta_0}{4}H_{\text{CFT}}}|b\rangle$, where $|b\rangle$ is a conformal
boundary state. $|b\rangle$ itself has no real space entanglement, and the correlation length is zero. 
By including the factor $e^{-\beta_0 H_{\text{CFT}}}$, a finite correlation length of order $\beta_0$ is introduced. 
In addition, it is found that the energy density of the system in $|\phi_0\rangle$ is the same as that in a thermal
ensemble with temperature $\beta^{-1}_0$, \textit{i.e.}, $\langle \phi_0|\mathcal{H}(x)|\phi_0\rangle=\text{Tr}(\mathcal{H}(x)e^{-\beta_0 H_{\text{CFT}}})$ \cite{Xueda2018}.
In this work, we assume $\beta_0\ll l$, such that the initial state can provide enough energy to `thermalize' the system 
after a quantum quench.

To study the entanglement Hamiltonian as well as the entanglement entropy for subsystem $A$, we consider the corresponding reduced density matrix
$\rho_A(t)=\text{Tr}_{\bar{A}}(|\psi(t)\rangle \langle \psi(t)|)
=\text{Tr}_{\bar{A}}(e^{-iH_{\text{CFT}}t-H_{\text{CFT}} (\beta_0/4) } |b\rangle \langle b| e^{iH_{\text{CFT}}t-H_{\text{CFT}} (\beta_0/4) }  )$,
where $\bar{A}$ denotes the complement of subsystem $A$. 
The path integral presentation of $\rho_A$ in Euclident spacetime ($\tau=it$) is shown in the following: 
\begin{eqnarray}\label{ConformalMapping}
\small
\begin{tikzpicture}[baseline={(current bounding box.center)}]

\draw [thick](0pt,20pt)--(100pt,20pt);
\draw [thick](0pt,-20pt)--(100pt,-20pt);
\draw [thick](0pt,-20pt)--(0pt,20pt);
\draw [thick][gray](0pt,5pt)--(40pt,5pt);
\draw (42pt,5pt) circle [radius=2pt];

\draw (95pt, 27pt)--(95pt,37pt);
\draw (95pt, 27pt)--(105pt,27pt);
\node at (101pt,32pt){$z$};

\node at (-12pt,20pt){$\beta_0/4$};
\node at (-14pt,-20pt){$-\beta_0/4$};

\node at (42pt,-2pt){$z_0$};

\end{tikzpicture}
\quad
\rightarrow
\quad
\begin{tikzpicture}[baseline={(current bounding box.center)}]

\draw [thick](0pt,0pt) ellipse (8pt and 20pt);
\draw [thick](80pt,0pt) ellipse (8pt and 20pt);
\draw (0pt,20pt)--(80pt,20pt);
\draw (0pt,-20pt)--(80pt,-20pt);

\draw [>=stealth,->] (105pt, 0pt)--(105pt,15pt);
\draw [>=stealth,->] (105pt, 0pt)--(120pt,0pt);
\node at (125pt,0pt){$u$};
\node at (110pt,15pt){$v$};

\draw (95pt, 27pt)--(95pt,37pt);
\draw (95pt, 27pt)--(105pt,27pt);
\node at (101pt,32pt){$w$};
\end{tikzpicture}
\end{eqnarray}
where $\rho_A$ is defined inside a semi-infinite rectangle on $z$-plane ($z=x+iy$), with conformal boundary condition $|b\rangle$ 
imposed along the boundary $x=0$ and $y=\pm \beta_0/4$. 
The branch cut (gray line) lies along $C=\{x+i\tau, 0\le x\le l\}$, and a small disc of radius $\epsilon$ has been removed at the entangling 
point $z_0=l+i\tau$ as a regularization.
One can impose conformal boundary condition $|a\rangle$ along the circle
centered at $z_0$.
Then one can consider the following conformal mapping
\be\label{ConformalMapfz}
w=f(z)=-\ln\left[
\frac{1+\sinh(2\pi(l-i\tau)/\beta_0)}{1+\sinh(2\pi(l+i\tau)/\beta_0)}\cdot\frac{\sinh(2\pi z/\beta_0)-\sinh(2\pi(l+i\tau)/\beta_0)}{\sinh(2\pi z/\beta_0)+\sinh(2\pi(l-i\tau)/\beta_0)}
\right]
\ee
to map the semi-infinite rectangle with a small disc removed at $z_0=l+i\tau$ to a cylinder in $w$-coordinate ($w=u+iv$), as shown in the right plot of \eqref{ConformalMapping}. 
The small circle at $z_0$ in $z$-plane is mapped to the right edge of the cylinder, and the boundary of the semi-infinite rectangle is mapped to the
left edge of the cylinder.
The cylinder is of circumference $2\pi$ and length $\text{Re}[f(l-\epsilon+i\tau)-f(i\tau)]$. 
Then for the entanglement Hamiltonian as defined through $\rho_A=e^{- H_E}$, one can find it is the generator of translation in $v$ direction
on the cylinder. Explicitly, one has
\be\label{HE_general}
H_E=-2\pi \int_{v=\text{const}} T_{vv}du=2\pi\int_{f(C)} T(w) dw+2\pi\int_{\overline{f(C)}} \overline{T}(\bar{w}) d\bar{w}
=2\pi\int_C \frac{T(z)}{f'(z)} dz +2\pi\int_{\bar{C}} \frac{\overline{T}(\bar{z})}{\overline{f'(z)}}d\bar{z}.
\ee
where $T$ and $\bar{T}$ are the holomorphic and anti-holomorphic components of energy momentum tensor.
They are related to the hamiltonian density $T_{00}$ and the momentum density $T_{10}$ (in Minkowski signature) as $T=(T_{00}+T_{10})/2$, and 
$\overline{T}=(T_{00}-T_{10})/2$.
Based on Eqs.\eqref{ConformalMapfz} and \eqref{HE_general}, one can find the \textit{exact} expression of entanglement Hamiltonian as 
\be\label{EH_exact}
\begin{split}
H_E(t)=&-2\beta_0 \int_0^l \frac{
\sinh[\frac{\pi(x-l)}{\beta_0}] \cosh[\frac{\pi (x-2t+l)}{\beta_0}] \sinh[\frac{\pi(x+l)}{\beta_0}] \cosh[\frac{\pi (x-2t-L)}{\beta_0}]
}{
\cosh(\frac{2\pi}{\beta_0}t)\,\sinh(\frac{2\pi}{\beta_0}L) \, \cosh[\frac{2\pi}{\beta_0}(x-t)]
} T(x,t) dx\\
&-
2\beta_0 \int_0^l \frac{
\sinh[\frac{\pi(x-l)}{\beta_0}] \cosh[\frac{\pi (x+2t+l)}{\beta_0}] \sinh[\frac{\pi(x+l)}{\beta_0}] \cosh[\frac{\pi (x+2t-L)}{\beta_0}]
}{
\cosh(\frac{2\pi}{\beta_0}t)\,\sinh(\frac{2\pi}{\beta_0}L) \, \cosh[\frac{2\pi}{\beta_0}(x+t)]
} \overline{T}(x,t) dx.
\end{split}
\ee 
There is much information contained in this exact form of entanglement hamiltonian. For example, at $t=0$, if we consider $\beta_0\gg l$
 (the correlation length is much larger than the typical system length), we obtain 
the entanglement Hamiltonian for the ground state of a CFT, \textit{i.e.}, $H_E\simeq 2\pi \int_0^l\frac{l^2-x^2}{2l} T_{00}(x)dx$ \cite{Tonni2016}. 
On the other hand, if 
$\beta_0\ll l$, we obtain the entanglement Hamiltonian for a short-range entangled state, 
$H_E\simeq \beta_0 \int_0^l \sinh[\frac{2\pi}{\beta_0}(l-x)]T_{00}(x)dx$.
For $t>0$, one can find that, in the limit $\beta_0\ll l$,
$H_E(t)$ in Eq.\eqref{EH_exact} can be simplified.
Remarkably, in the long time limit $t\gg l$, $H_E(t)$ is exactly the same as that in a thermal ensemble form, with the expression
\be
H_E=2\beta_0 \int_0^l\frac{
\sinh[\frac{\pi (l-x)}{\beta_0}]\,\sinh[\frac{\pi(l+x)}{\beta_0}]
}{
\sinh(\frac{2\pi l}{\beta_0})
}T_{00}(x)dx.
\ee 
This is the Eq. \ref{eq:cft2} shown in the main text. 
Please note that, $T_{00}(x)$ is the hamiltonian density of the CFT Hamiltonian:
$H_{\text{CFT}}= \int dx T_{00}(x) $. 
To gain a general picture about this result, in Fig. \ref{sfig:cft}, we plot the spatial dependence of $\beta(x)/\beta_0$ for various temperature $\beta_0$,
where the envelop function is $\beta(x)=2\beta_0 \frac{
	\sinh[\frac{\pi (l-x)}{\beta_0}]\,\sinh[\frac{\pi(l+x)}{\beta_0}]}{	\sinh(\frac{2\pi l}{\beta_0})}$.
As we can see, 
the envelope function is almost flat in $|x-\ell|>\mathcal{O}(\beta_0)$. 
The higher temperature (smaller $\beta_0$) we set, 
a sharper change close to the entanglement cut $x\sim \ell$ will be obtained.

\begin{figure}
	\includegraphics[width=0.40\textwidth]{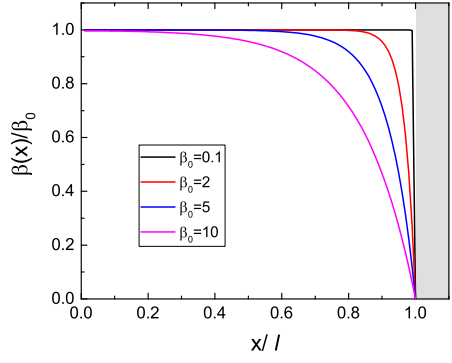}	
	\caption{Spatial dependence of the envelope function $\beta(x)/\beta_0$. Here we choose $\ell =10$.
	} \label{sfig:cft}
\end{figure}

\section{Entanglement entropy evolution after a global quantum quench}

Based on the setup in the previous section, it is straightforward to evaluate the time evolution of entanglement entropy as follows.
We first consider the Renyi entropy $S^{(n)}_A $.
Then the von Neumann entropy can be obtained by taking the limit $S_A=\lim_{n\to 1} S^{(n)}_A$.
The Renyi entropy is defined as
\be\label{Sn_def}
S^{(n)}_A :=\frac{1}{1-n}\ln \frac{\text{Tr}(\rho_A^n)}{(\text{Tr} \rho_A)^n}=\frac{1}{1-n}\ln \frac{Z_n}{(Z_1)^n},
\ee
where the partition function $Z_n$ can be obtained by gluing $n$ copies of cylinders in \eqref{ConformalMapping} along the branch cuts. That is, 
$Z_n$ is defined on a cylinder of circumference $2n\pi$ and length $W=\text{Re}[f(l-\epsilon+i\tau)-f(i\tau)]$.
To evaluate $Z_n$, instead of considering a Hamiltonian evolving in $v$ direction on the cylinder, 
now we consider the Hamiltonian evolving along $u$ direction. 
Then we have
\be
Z_n=\langle b|e^{-H_{\text{CFT}} \cdot W}|a\rangle=
\langle b|e^{-\frac{2\pi}{2\pi n} (L_0+\overline{L}_0-\frac{c}{12})\cdot W}|a\rangle
=\sum_{n,n'}
\langle b|n\rangle \langle n|e^{-\frac{2\pi}{2\pi n} (L_0+\overline{L}_0-\frac{c}{12})\cdot W}|n'\rangle \langle n'|a\rangle
\ee
where the sum is over all allowed bulk operators, and $|b\rangle$ ($|a\rangle$) denotes the conformal boundary state defined on the left (right)
edge of the cylinder in \eqref{ConformalMapping}. In the limit $\frac{W}{2\pi n}\gg 1$, which is the case we considered here, only the ground state
$|0\rangle$ dominates, and $Z_n$ can be simplified as
\be\label{Zn}
Z_n\simeq e^{\frac{c}{12 n} W}\cdot \langle b|0\rangle\cdot \langle 0|a\rangle.
\ee
Then based on Eqs.\eqref{Sn_def} and \eqref{Zn}, we can obtain
\be
S_A^{(n)}=\frac{c}{12}\cdot \frac{1+n}{n}\cdot W-g_a-g_b,
\ee
where $g_{a,b}=-\ln \langle a,b|0\rangle$ is the so-called Affleck-Ludwig boundary entorpy. The length of cylinder $W$ can be evaluated 
based on the conformal mapping in Eq.\eqref{ConformalMapfz}. After some straightforward algebra, one can find that $W$ depends on time $t$ 
as follows:
\be\label{W}
W\simeq \left\{
\begin{split}
&\ln\frac{\beta_0}{2\pi \epsilon}+\frac{2\pi}{\beta_0}t,\quad t<l,\\
&\ln\frac{\beta_0}{2\pi \epsilon}+\frac{2\pi}{\beta_0}L,\quad t>l.
\end{split}
\right.
\ee
In a lattice model, the UV cutoff $\epsilon$ can be considered as the lattice constant. Since the initial state is short-range correlated ($\beta\ll l$),
the first term in \eqref{W} is $\mathcal{O}(1)$. Then the leading terms of Renyi entropy and von-Neumann entropy are
\be
S_A^{(n)}(t)\simeq
\left\{
\begin{split}
&\frac{\pi c}{6\beta_0}\cdot \frac{1+n}{n}\cdot t, \quad t<l\\
&\frac{\pi c}{6\beta_0}\cdot \frac{1+n}{n}\cdot l, \quad t>l.
\end{split}
\right.
\quad\quad
S_A(t)\simeq
\left\{
\begin{split}
&\frac{\pi c}{3\beta_0}\cdot  t, \quad t<l\\
&\frac{\pi c}{3\beta_0}\cdot l, \quad t>l.
\end{split}
\right.
\ee


\section{Additional results on the entanglement Hamiltonian}
\subsection{Consistency on different system sizes}
In the main text, we show the time evolution of the entanglement Hamiltonian. 
The results in the main text is for a given total system size $L=48$ and subsystem length $\ell=9$.
Actually, we have checked different system sizes and confirmed that the features shown in this work are robust against the finite-size effects.
Next we show that the entanglement Hamiltonian on the different system sizes. 

In Fig. \ref{sfig:EH_time}, we show the time evolution of the entanglement Hamiltonian for a larger total system size with $L=60$. 
The subsystem length is set to be $\ell=9$. 
By comparison different system sizes, we find the very similar features: 
The local coupling strengths fluctuates in the short time regime, while 
in the long-time limit the local coupling strengths approach nearly stationary values.
The phase fluctuations of boson hopping are nonzero in the short time regime and tend to vanish in the long time limit.
The consistency reaching on different system sizes show that, 
the general features that we discovered is robust, 
which is not finite size effects.

\begin{figure*}[t]
	\includegraphics[width=0.83\textwidth]{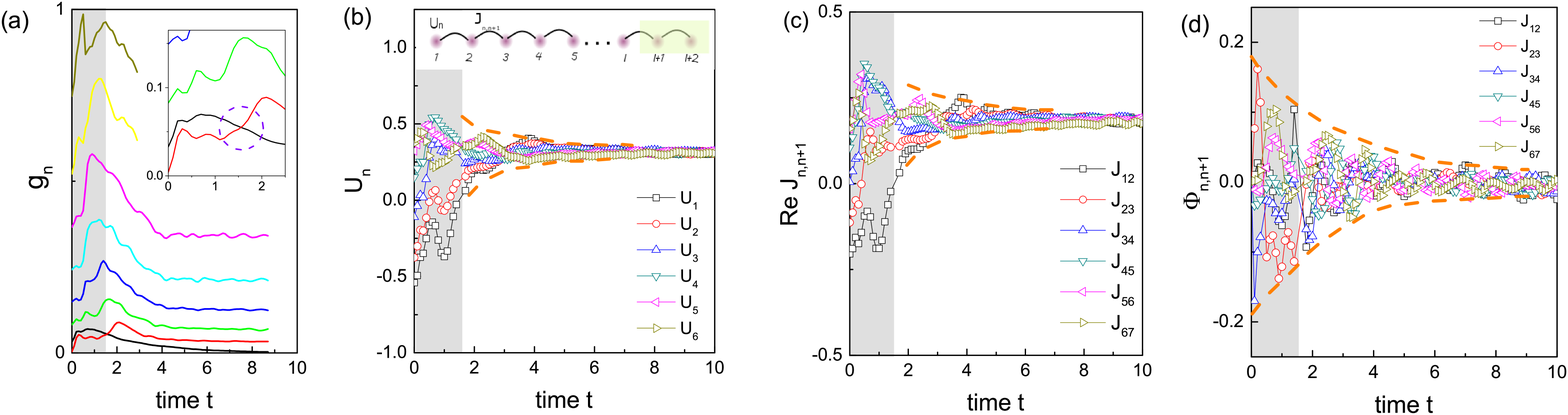}
    \includegraphics[width=0.83\textwidth]{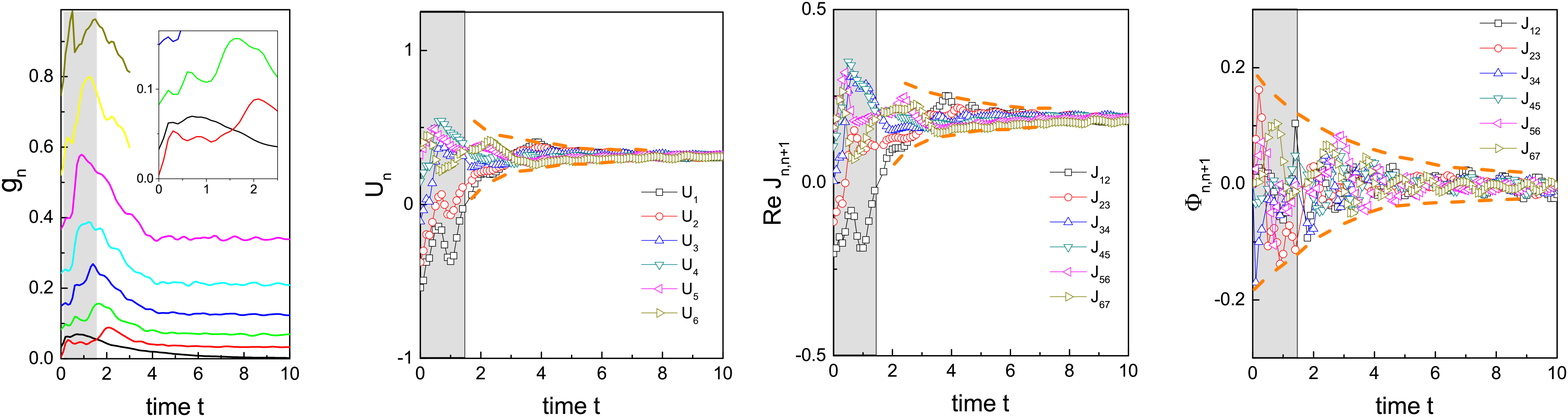}
	\caption{Time evolution of the EH on different total system sizes: (top) $L=48$ and (bottom) $L=60$.
		The parameters of the EH as a function of time:
		interaction strength $U_n$, real part of couplings $ Re J_{n,n+1}(t)$, phase part of couplings $\Phi_{n,n+1}=\arg J_{n,n+1}(t)$. 
		Here we quench the Bose-Hubbard model from $U^{\mathbf{in}}=5.0$ to $U^{\mathbf{fi}}=3.3$.
		Different symbols label local coupling and interaction strengths.
		Here we set the typical subsystem length is $\ell=9$.  
	} \label{sfig:EH_time}
\end{figure*}

We also checked that the subsystem length $\ell$ doesnot change the main features as we discussed in the main text.
In Fig. \ref{sfig:EH_BH_scaling}, we compare the spatially distributed interaction strength for different subsystem length $\ell$. 
The interaction strength is almost uniformly distributed when the position is away from the entanglement cut $x\ll \ell$,
while it experiences a reduce when approaching the entanglement cut position $x\sim \ell$.
Clearly, it shows the scaling behavior from the CFT works so good for all subsystem length $\ell$. 
Thus, all features of the entanglement Hamiltonian are robust when we tune subsystem length $\ell$ (we need to force $\ell \ll L$).

\begin{figure}	
	\includegraphics[width=0.7\textwidth]{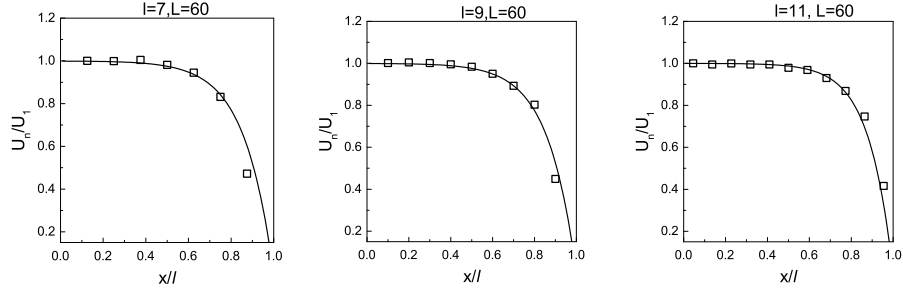}
	\caption{The EH (at long time limit ) of different subsystem sizes $\ell=7,9,11$. 
		The total system size is set to be $L=60$.
		Spatial dependence of local coupling strengths for quenching parameters:
		$U^{\mathbf i}=5.0,U^{\mathbf f}=3.3$ (black squares). 
		The solid lines show best fit to the envelope function Eq. (\ref{eq:cft2}).
	} \label{sfig:EH_BH_scaling}
\end{figure}

\begin{figure}[!htb]
	\includegraphics[width=0.53\textwidth]{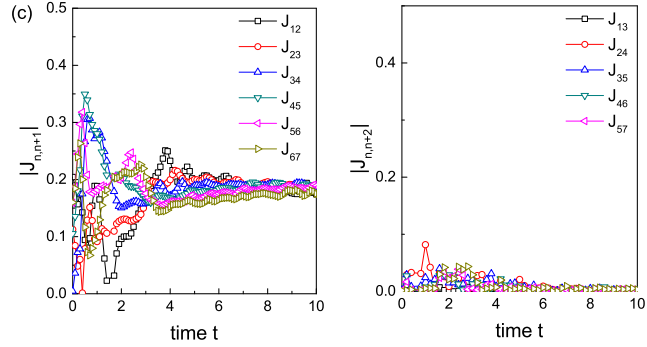}
	\caption{(left) Time-evolution of the hopping strength of nearest-neighbor terms $|J_{n,n+1}|$
		(right) Time-evolution of the hopping strength of second nearest-neighbor terms $|J_{n,n+2}|$.
		The subsystem length is $\ell=9$ and total system length is $L=48$.
	} \label{sfig:EH_BH_nn}
\end{figure}

\subsection{Long-ranged hopping strength in the entanglement Hamiltonian}
In the main text, we only show that the entanglement Hamiltonian with nearest-neighbor hopping terms 
and on-site Hubbard interactions. One may wonder how the long-ranged couplings could influence the entanglement Hamiltonian. 
Here we add second nearest-neighbor hopping terms, $J_{n,n+2} b^\dagger_{n}b_{n+2} + J^*_{n,n+2}b_{n}b^\dagger_{n+2}$, 
to the trial entanglement Hamiltonian.
The calculations are parallel to that in the main text. 
In Fig. \ref{sfig:EH_BH_nn}, we show the time evolution of the second neighbor couplings $|J_{n,n+2}(t)|$.
By comparison with the first neighbor couplings $|J_{n,n+1}(t)|$, we found that the second neighbor couplings 
are much smaller in the short time limit. 
More importantly, in the long time limit, the second neighbor couplings $|J_{n,n+2}(t)|$ converges to zero with little fluctuations,
which means they identically vanishes in the entanglement Hamiltonians,
in contrast to the first neighbor couplings (which converges to nonzero values). 
Here we conclude that the long-time entanglement hamiltonian doesnot contain the long-ranged couplings, 
consistent with the conformal field theory prediction.

\subsection{Frequency analysis of entanglement Hamiltonian}
In the main text, we elucidate that the entanglement Hamiltonian approaches stationary in the long time limit. 
Here we provide further analysis to support it. 
In Fig. \ref{fig:EH_fft}, we show the fourier transformation of the long-time entanglement Hamiltonian: $F(\omega)=\int dt F(t) e^{i\omega t}$. 
It is clear to see that, zero frequency mode dominates for real part of coupling strength and interaction strength,
where zero frequency mode in $Re J(\omega=0)$ and $U(\omega=0)$ are at least two orders larger than the other frequencies.
For the imaginary part of coupling strength, despite of fluctuations around zero (as shown in the main text), 
 zero frequency channel ($Im J(\omega=0)$) is still larger than other nonzero frequencies. 
Physically, this frequency analysis indicates that, in the real-time domain,
the entanglement Hamiltonian is nearly stationary in the long-time limit.

\begin{figure}[!htb]
	\includegraphics[width=0.73\textwidth]{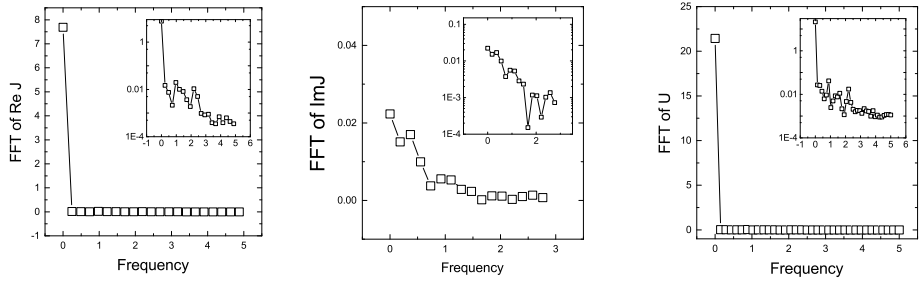}
	\caption{Fourier transformation of long-time entanglement Hamiltonian for 
		(a) real part of coupling strength $Re J$,
		(b) imaginary part of coupling strength $Im J$ 
		and (c) interaction strength $U$. 
		The system size is $L=36$ and $\ell=9$. We only choose the results in the long-time limit and make fourier transformation.
	} \label{fig:EH_fft}
\end{figure}

\subsection{Quantum dynamics of entanglement Hamiltonian on spin$-1/2$ XXZ model}
In the main text, we numerically map out the dynamics of entanglement Hamiltonian $H_E$
in Boson Hubbard model. The main conclusion has been drawn based on the results in Bose-Hubbard model,
which has been a widely implemented in the cold atom experiments. 
Next, to demonstrate that the phenomenon shown in the main text is not dependent on specific models,
in this section we will apply the numerical scheme to study a spin$-1/2$ XXZ model.
We investigate the one dimensional spin$-1/2$ XXZ model:
\begin{eqnarray}\label{eq:Hei}
\hat H= J^{xy} \sum_i (S^x_i S^x_{i+1} +S^y_{i} S^y_{i+1})+ J^{zz} \sum_{i} S^z_i S^z_{i+1} .
\end{eqnarray}
In the equilibrium case, this model hosts a critical spin liquid state as the ground state for 
$-1<J^{zz}\leq1$ and Ising-like magnetic order for $1<J^{zz}$ (by setting $J^{xy}=1$).
We will study the quantum dynamics by suddenly changing the parameters in Hamiltonian Eq. \ref{eq:Hei}.
We focus on a quantum quench from $J^{zz}>1$ (Ising-like magnetic order) to $J^{zz}=1$ (critical liquid).
Once the $|\Psi(t)\rangle$  is computed, we partition a one-dimensional chain with length $L$ into two segments, $\ell$ and $L-\ell$,
and calculate the reduced density matrix of the $A$, $\rho_{A}(t)=Tr_{B} |\Psi(t)\rangle \langle \Psi(t)|$.
The von Neumann entanglement entropy is $S_{\ell}(t)=-\sum_i \lambda_i(t) \ln\lambda_i(t)$,
where $\lambda_i$ are the eigenvalues of $\rho_A(t)$.
The entanglement Hamiltonian is defined by $\rho_A(t)=\exp(-\hat H_E)$.
The operator form of the entanglement Hamiltonian is obtained using the numerical scheme that is introduced in the main text.
The bond dimension up to $2048$ guarantees the neglected weight in the Schmidt decomposition in  each  time step is less than $10^{-6}$.

In Fig. \ref{fig:EH_Hei}(a) we show the entanglement entropy dependence on time. 
The total system size $L=48$, and we confirmed the physics here doesnot change when we increase total system size to $L=80$.
For all subsystem length $\ell$, $S_{\ell}(t)$ shows three time regimes:
a linear increasing regime for $t<t_*$, a non-linear increasing regime for $t \gtrsim t_*$ and
a saturation regime in the long time limit $t\gg t_*$.
At early time $t<t_*$, the entanglement entropy grows
linearly with time due to the ``ballistic'' propagation of entanglement.
At intermediate time $t>t_*$, $S_{\ell}(t)$ slowly increases with the time.
At the long-time $ t\gg t_*$, $S_{\ell}(t)$ saturates to its
steady-state value.
This saturation begins earlier for smaller $\ell$.
In Fig. \ref{fig:EH_Hei}(b), the equilibrium value of entanglement entropy at the long-time linearly depends on subsystem size $\ell$,
which clearly exhibits a ``volume-law'' scaling.

The evolution of the entanglement Hamiltonian is shown in Fig. \ref{fig:EH_Hei}(c-e).
The related entanglement Hamiltonian is defined by
$H_E = \sum_n J^{zz}_{n,n+1}(t) S^z_n S^z_{n+1} + J^{xy}_{n,n+1} (t)(S^x_n S^x_{n+1} +S^y_n S^y_{n+1})$.
In general, the coupling parameters of $H_E$ shows strong oscillations in the short time regime.
At intermediate regime $t\lesssim t_*$, the imaginary part $Im J^{xy}_{n,n+1}$ is nonzero, indicating 
an emergent current flow in the process towards equilibration.
In particular, at the long-time $t \gg t_*$, the local coupling strengths approach steady-state values and doesnot change with time.
In the steady-state, the local coupling strengths show a almost uniform distribution in spatial space (indicated by red dashed line).
We find that the above features are the same as those in Bose-Hubbard model. 
Based on this, we conclude that
the main findings in the paper is robust and general, not specific to models.

\begin{figure}[!htb]
	\includegraphics[width=0.65\textwidth]{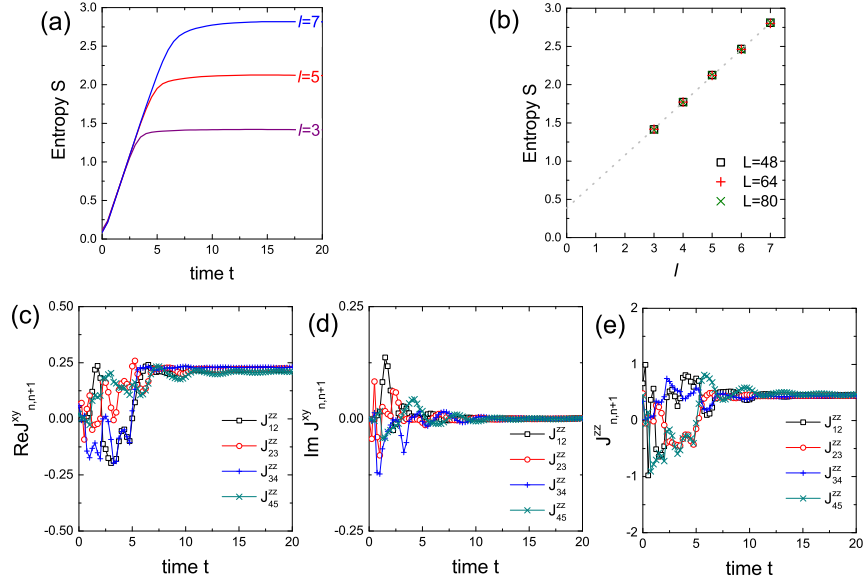}\\
	\caption{Quantum dynamics of spin-$1/2$ XXZ model by quenching from $J^{zz}=4.0$ to $J^{zz}=1.0$.
		(a) Time evolution of entanglement entropy for various subsystem length $\ell$. The total system size is $L=48$.
		(b) Long-time entanglement entropy versus subsystem length $\ell$. The grey dashed line is the linear fit.
		The total system size is $L=48$ (black square), $L=64$ (red cross) and $L=84$ (green cross).
		(c-e) Time evolution of local coupling strength $J^{xy,zz}(n,n+1)(t)$ of the entanglement Hamiltonian:
		$H_E = \sum_n J^{zz}_{n,n+1}(t) S^z_n S^z_{n+1} + J^{xy}_n (t)(S^x_n S^x_{n+1} +S^y_n S^y_{n+1})$.
		The subsystem size is chosen to be $\ell=6$.
	} \label{fig:EH_Hei}
\end{figure}


\end{widetext}

\end{document}